\documentclass[aps,prl,twocolumn,groupedaddress]{revtex4}
\usepackage{graphicx} 
\usepackage{multirow}
\usepackage{amsmath}
\begin{document}


\title{Hidden Fermi Liquid: Self-Consistent Theory for \\
the Normal State of High-T$_{c}$ Superconductors}
\author{Philip A. Casey}
\author{Philip W. Anderson}
\affiliation{Princeton University, Department of Physics, Princeton, New Jersey 08544, USA}
\date{\today}

\begin{abstract}
Hidden Fermi liquid theory explicitly accounts for the effects of Gutzwiller projection in the t-J Hamiltonian, widely believed to contain the essential physics of the high-T$_{c}$ superconductors. We derive expressions for the entire ``strange metal'', normal state relating angle-resolved photoemission, resistivity, Hall angle, and by generalizing the formalism to include the Fermi surface topology - angle-dependent magnetoresistance. We show this theory to be the first self-consistent description for the normal state of the cuprates based on transparent, fundamental assumptions.  Our well-defined formalism also serves as a guide for further experimental confirmation.
\end{abstract}
\maketitle
The anomalous ``strange metal'' properties of the normal, non-superconducting state of the high-T$_{c}$ cuprate superconductors have been extensively studied for over two decades.\cite{Anderson1997, Varma1989, Orenstein2000, Shen1995, Damascelli2003, Hussey2008} The resistivity is robustly T-linear at high temperatures while at low T it appears linear near optimal doping and is T$^{2}$ at higher doping. The inverse Hall angle is strictly T$^{2}$ and hence has a distinct scattering lifetime from the resistivity. The transport scattering lifetime is highly anisotropic as directly measured by angle-dependent magnetoresistance (ADMR, or similarly AMRO)\cite{Hussey2003, Abdel-Jawad2006, Analytis2007, French2009} and indirectly in more traditional transport experiments. The IR conductivity exhibits a non-integer power-law in frequency\cite{Schlesinger1990, vanderMarel2003}, which we take as a defining characteristic of the ``strange metal''. 

A phenomenological theory of the transport and spectroscopic properties at a self-consistent and predictive level has been much sought after, yet elusive. We demonstrate here that the hidden Fermi liquid theory (HFL)\cite{Anderson2008, Casey2008, Anderson2009, Anderson2009b, Anderson2010, Casey2010} is the effective low-energy theory for the normal state and no longer just a proposal. After reviewing the theory, we derive well-defined expressions relating ARPES, resistivity, Hall angle, and ADMR. Self-consistency is shown in the one system where most datasets are available, overdoped (OD) $Tl_{2}Ba_{2}CuO_{6+\delta}$. 

In the cuprate phase diagram, a line is usually drawn up and to the right from the edge of the superconducting dome in the OD region to separate the ``strange metal'' from a conventional Fermi liquid (FL) at high doping. We argue there is no crossover from the ``strange metal'' for high doping. HFL is valid for the entire normal state and we will show that the data only appears FL-like as the HFL bandwidth, W$_{HFL}$, becomes large at high doping. 

Some proposals, which have neither demonstrated self-consistency nor explained the IR conductivity, invoke the heuristic that there is single scattering lifetime in the problem with ``hot'' or ``cold'' spots\cite{Hlubina1995, Zheleznyak1998, Ioffe1998} with different temperature dependences in different regions of the Fermi surface. 
\begingroup
\squeezetable
\begin{table}[hpb]
\renewcommand{\baselinestretch}{.1}
\renewcommand{\arraystretch}{.1}
\begin{tabular}{|c|c|}
\hline 
{\normalsize ``Quasiparticles'' $\hat{c}^{\dagger}_{i \sigma}$,$\hat{c}_{i \sigma}$} & {\normalsize ``Pseudoparticles'' $c^{\dagger}_{i \sigma}$,$c_{i \sigma}$} \\
\hline
Physical electrons of the system & True excitations of the system \\
\hline
Excitations of the (reduced & Excitations of the unprojected \\
dimension) projected space & space, the HFL \\
\hline
Automatically enforce the projection, & HFL ansatz states that these obey \\ 
lend no spectral weight to UHB & conventional Fermi liquid rules \\
\hline
Accelerated by an electric field, & Magnetic field commutes with \\
act as a decay bottleneck & projection, so Hall effect, dHvA, \\
in the resistivity & etc. will be that of the HFL \\
\hline
Do not transfer momentum to the & Transfer momentum to the lattice \\
lattice but rather decay into the & via umklapp scattering, \\
HFL, $\Gamma_{decay} = p \pi T$ & $\Gamma_{HFL} = T^{2}/(2 W_{HFL})$ \\
\hline 
\multicolumn{2}{|c|}{Renormalized Fermi velocity, $v_{F,ren} = v_{F,0} g_{t}(x)$} \\
\multicolumn{2}{|c|}{Anisotropy in $v_{F}$, $k_{F}$ given by band structure} \\
\hline 
\end{tabular}
\caption{Comparison of the two distinct excitations of the HFL theory - the quasiparticles and the HFL pseudoparticles.}
\end{table}
\endgroup
However, the more recent experimental probe of ADMR allows a direct test of the anisotropic scattering predicted by any theory. We show that the anisotropy observed in the transport scattering rate in the ADMR is precisely reproduced by including the slight variation of $v_{F}$ and $k_{F}$ around the Fermi surface within HFL.

All data is described using only one main assumption: that the Hubbard energy, U, of excitations into doubly occupied Wannier orbitals on a particular site is large. Antibound states, representing real double occupancy, are thus prohibited. This assumption requires us to use the Gros-Rice canonical transformation of the more general Hubbard model to renormalize U to infinity\cite{Rice1988} and the HFL formalism starts with the t-J Hamiltonian,
\begin{subequations}
\begin{eqnarray}
&\displaystyle H_{t-J} = P[ \sum_{i,j, \sigma} t_{ij} c_{i \sigma}^{\dagger} c_{j \sigma}]P + \sum_{i,j} J_{ij} S_{i} \cdot S_{j}& \\
&\displaystyle P = \prod_{i} (1-n_{i\uparrow}n_{i\downarrow})&
\end{eqnarray}
\end{subequations} 
where P is the Gutzwiller projector that will enforce the exclusion of antibound states. In the normal state, above T$^{*}$, the temperature is sufficient to break the superconducting pairs caused by the exchange, J. The system is thus gapless and J only enters into the calculation as an effective interaction that renormalizes the kinetic energy in the Shankar sense - we have a HFL and not a hidden Fermi gas. We can write the effective Hamiltonian for the normal state as a projected kinetic energy,\cite{Anderson2006}
\begin{equation}
\displaystyle H_{proj} = \sum_{i,j, \sigma} t_{ij} P c_{i \sigma}^{\dagger} c_{j \sigma} P = \sum_{i,j, \sigma} t_{ij} \hat{c}_{i \sigma}^{\dagger} \hat{c}_{j \sigma}
\end{equation}
We introduce the projective quasiparticles (QPs), denoted with ``hats'', which enforce the projection and have no amplitude in the upper Hubbard band (UHB). These excitations span the projected Hilbert space of reduced dimension. These are the physical electrons that a photoemission experiment would remove, for example. We can expand these in terms of three pseudoparticle operators (Eq. (3), see Table I), denoted as ``hatless''. The HFL ansatz,\cite{Anderson2009} justified by our self-consistency with experiment, states that these obey conventional FL rules - for example, they have a non-zero residue, Z.
\begin{subequations}
\begin{eqnarray}
\displaystyle \hat{c}_{i \sigma} = c_{i \sigma} P = c_{i \sigma} c_{i - \sigma} c_{i - \sigma}^{\dagger} \\
\displaystyle \hat{c}_{i \sigma}^{\dagger} = P c_{i \sigma}^{\dagger} = c_{i - \sigma} c_{i - \sigma}^{\dagger} c_{i \sigma}^{\dagger}
\end{eqnarray}
\end{subequations} 

The pseudoparticles (PPs) span the unprojected Hilbert space of many-body wavefunctions. Individually they do not represent the physical electrons, rather they are the true excitations of the system. To see this, we start by noting that the solution for the wavefunction of the problem is projective in nature, $\Psi (r_{1}, r_{2}, \ldots, r_{N}) = P \Phi (r_{1}, r_{2}, \ldots, r_{N})$ - where $\Phi$ is the completely general many-body wavefunction. We can write the Schrodinger equation as, $H_{t-J} \Psi = \epsilon \Psi$. But to represent the excitations of the system at their most fundamental level is not to consider this to be an equation for $\Psi$ but to observe that because P$^{2}$ = P and $\epsilon$ commutes with P, the Schrodinger equation can be re-written, $H_{t-J} \Phi =P (\epsilon \Phi)$. Indeed P acts to the right on $(\epsilon \Phi)$, but regardless the true excitations are fundamentally of the unprojected, $\Phi$-space - these are the HFL PPs. This equation is undercomplete and does not have a unique solution. In the normal state however, we can choose the ground state, $\Phi_{0}$, to be the Hartree-Fock product form of a Fermi sea. The resulting mean-field Hartree-Fock equations give the energy spectrum of the PPs and momentum conservation then makes it unique.\cite{Anderson2006, Anderson2008} The QPs of the system are then given using Eq. (3), but there is no adiabatic continuation between the QPs and PPs - as is the basis in FL theory of relating the ``dressed'' excitations back to the ``bare'' ones. The metric of this correspondence, Z, vanishes and the physical space is a non-Fermi Liquid.

The Green's function of the QPs was calculated, explicitly accounting for projection.\cite{Casey2008, Anderson2006} The result is a product of two terms; the first is the conventional FL term that would result in a Lorentzian spectral function. 
\begin{equation}
\displaystyle G_{proj, i \sigma}^{holes}=\langle0|\hat{c}_{i \sigma}^{\dagger}(t) \hat{c}_{i \sigma}(0)|0\rangle\cong\langle0|c_{i \sigma}^{\dagger}(t) c_{i \sigma}(0)|0\rangle G_{i \sigma}^{*}(t)
\end{equation}
The correction, $\displaystyle G_{i \sigma}^{*}(t)=\langle0| (1 - n_{i - \sigma})(t) (1 - n_{i - \sigma})(0) |0\rangle$, represents a canonical inverse X-ray edge problem if we treat the Gutzwiller projection at a singly-occupied site as a infinite, repulsive potential. $G^{*}_{i \sigma}(t)$ acts to project out a -$\sigma$ electron at t = 0, then measures the overlap of this decaying state with the originally projected state at time, t. This problem has been solved in a variety of ways, originally in ref. \cite{Nozieres1969}. But the most physically insightful solution is in terms of Tomonaga bosons\cite{Schotte1969} - the solution being that the projected state decays as a power-law, $G^{*}_{i \sigma}(t) \propto t^{-p}$. Using Nozieres' theorem and Friedel's sum rule, the doping dependent exponent was found to be $p=(1-x)^{2}$/4. This doping dependence is in close agreement over a wide doping range with the power-law dependence of the IR conductivity.\cite{Casey2008} This power-law decay extends to finite temperatures, replacing $t^{-p}$ by $e^{-p \pi Tt}$ (for $Tt \gg$ 1)\cite{Yuval1970} and we define $\Gamma_{decay}$ = p$\pi$T.

We compared this projective Green's function with the nodal laser-ARPES in $Bi_{2}Sr_{2}CaCu_{2}O_{8+\delta}$\cite{Koralek2006} and found excellent agreement.\cite{Casey2008} The predicted $\omega^{-1+p}$ decay of the EDC's accounted for the significant spectral weight at high energies, historically mistaken as experimental background. The ARPES scattering rate was extracted as a sum of two rates, C$_{HFL}$ the only free parameter,
\begin{subequations}
\begin{eqnarray}
&\displaystyle \Gamma_{ARPES} = p \pi T + C_{HFL} v_{F}^{2} (k-k_{F})^{2}& \\
&\displaystyle W_{HFL} = \frac{3}{2 \pi^{2} k_{B} C_{HFL}}&
\end{eqnarray}
\end{subequations}

We now describe these two scattering rates in the context of transport. The $(k-k_{F})^{2}$ rate is the umklapp scattering that transfers momentum of the PPs in the unprojected space to the lattice, $\tau_{HFL}^{-1} = T^{2}/W_{HFL}$. W$_{HFL}$ is about an order of magnitude smaller than $\epsilon_{F}$ ($v_{F,ren}=v_{F,0}g_{t}(x)$ where $g_{t}(x)=2x/(1+x)$).\cite{Anderson2009} We calculate the T$^{2}$ expectation value of the $(k-k_{F})^{2}$ rate by weighting it by the density of states, given by the Fermi function derivative as PPs obey Fermi-Dirac statistics. We thus relate W$_{HFL}$ (Eq. (5b)) to $\Gamma_{ARPES}$. This does not coincide with FL theory, nor should it since Z=0. The lack of T$^{2}$ term in $\Gamma_{ARPES}$ is due to PP ``protection'' in the unphysical space. The non-FL QPs, which are directly probed in ARPES, are entangled and must break up into disentangled PPs prior to umklapp scattering.
\begin{figure}[hpb]
\includegraphics[scale=.325,clip=true, trim=14mm 9mm 8mm 26mm]{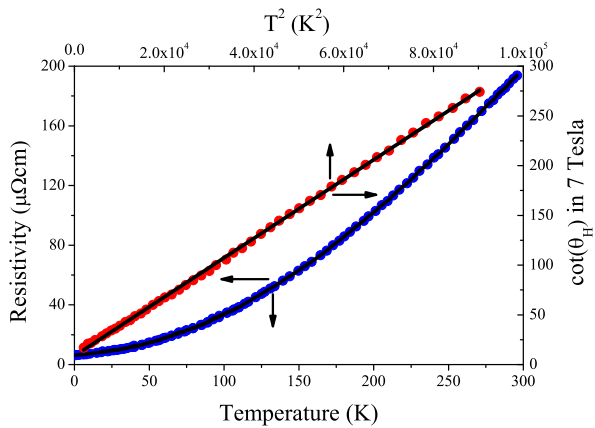}
\caption{Self-consistent HFL fits (black) of $\rho$ versus T and inverse Hall angle versus T$^{2}$. The sample is OD $Tl_{2}Ba_{2}CuO_{6+\delta}$ (x=0.26). The resistivity is given by Eq. (6) (pre-factor is a free parameter) with W$_{HFL}$ = 800 K. This linear best fit for $\cot \theta_{H}$ was used to determine $\alpha$ for Eq. (7), resulting in W$_{HFL}$ = 800 K. Data for the inverse Hall angle (red) and resistivity (blue) were extracted from ref. \cite{Mackenzie1996}. The magnetic field of 7 T in the Hall experiment is not expected to fully suppress superconductivity at the lowest temperatures so that portion of the Hall data should be ignored while the 16 T used in the $\rho$(T) experiment should be sufficient. (Color online)}
\end{figure}

Applying an electric field will excite QPs of the projected space, which will consequently decay into the HFL (with rate $\Gamma_{decay}$, above). Physically the decay process corresponds to the QP decaying into one PP and coupling to the Tomonaga bosons in the unprojected space.\cite{Anderson2009} The $G^{*}_{i \sigma}(t)$ correction previously discussed results in a predicted zero-bias anomaly in the tunneling density of states,\cite{Anderson2006} which broadens as $\Gamma_{decay}$, so the number of states available for the QPs to decay into increases linearly with T.\cite{Anderson2010} This decay process acts as a bottleneck that must occur before momentum can be transferred to the lattice by the PPs. When calculating the transport lifetime, $\tau_{tr}$, we must therefore add the decay lifetime with the HFL scattering lifetime, $\tau_{tr} = \tau_{decay} + \tau_{HFL}$.\cite{Anderson2009} The resistivity is thus given in terms of W$_{HFL}$ as,
\begin{equation}
\displaystyle \rho(T) = \frac{2 \pi^{2} \hbar p}{e^{2} \epsilon_{F}} \frac{T^{2}}{T + 2 \pi p W_{HFL}} + \rho_{res}
\end{equation} 
Fitting Eq. (6) to $\rho$(T) for OD $Tl_{2}Ba_{2}CuO_{6+\delta}$, yields W$_{HFL}$ = 800 K (see Fig. 1).

Applying a magnetic field does not excite projective QPs, which would decay into the HFL, rather only causes their trajectories to deviate. Hence, the magnetic field commutes with P and its effect acts directly on the unprojected space, $\Phi$. The Hall scattering lifetime is hence that of the HFL, $\tau_{H} = \tau_{HFL}$.\cite{Anderson2009} Since the inverse Hall angle is given by $\cot \theta_{H} = (\omega_{c} \tau_{H})^{-1}$ and $\tau_{HFL}^{-1} = T^{2}/W_{HFL}$, we can deduce W$_{HFL}$ from the Hall angle. Ref. \cite{Chien1991} derived a bandwidth in another context but we can follow their arguments for estimating the cyclotron frequency, $\omega_{c}$, leading to a relation between W$_{HFL}$ and $\cot \theta_{H}$,
\begin{equation}
\displaystyle W_{HFL} = \sqrt{\frac{n \pi \hbar}{\alpha e B}}
\end{equation} 
with n = $k_{F}^{2}/2 \pi$ being the carrier concentration and $\alpha$ the slope of $\cot \theta_{H}$ versus T$^{2}$. We estimate n as one carrier per unit cell, consistent with $k_{F}$ from both ARPES and ADMR at slightly different doping.\cite{Plate2005, Hussey2003} Using the $\cot \theta_{H}$ data\cite{Mackenzie1996} on $Tl_{2}Ba_{2}CuO_{6+\delta}$ (x = 0.26), yields W$_{HFL}$ = 800 K (Fig. 1), in excellent agreement with W$_{HFL}$ determined from the resistivity. Even higher precision of W$_{HFL}$ could have been achieved by accounting for the variation of $k_{F}$ and $v_{F}$ about the Fermi surface. To balance precision with transparency in our solution however, we invoke the fact that the bulk transport properties will be dominant along ($\pi, \pi$), where $v_{F}$ is maximal.

QPs may impurity scatter prior to decay, leading to $\rho_{res}$ and, less obviously, the T=0 offset in $\cot \theta_{H}$. As for the latter, there is no intrinsic contribution of the QPs to $\cot \theta_{H}$ as the magnetic field neither creates QPs nor do they transfer momentum via umklapp. But in the impurity scattering channel, QP momentum is transferred. The magnetic field deviates the trajectory of both PPs and QPs, so QPs will contribute to $\cot \theta_{H}$ as an impurity term. Impurity scattering is not a many-body effect and does not impact the unphysical PPs. It is a property of the geometry of the transmission channels and not a true dissipative process, a result of the Landauer formula for conductivity, and is unaffected by the bottleneck. 

An electron-phonon (e-ph) interaction exists with both PPs and QPs but it is too weak for umklapp scattering, evidenced by $\rho$(T) as there is neither high-T saturation nor indication of the Debye frequency. A temperature gradient thermally skews the Fermi surface (FS) so small q, normal e-ph scattering can inelastically thermally relax entropy carriers. The FS is not displaced so the gradient will not disrupt the equilibrium between QPs and PPs and the entropy transport is just that of the HFL.\cite{Casey2010}
\begin{figure}[hpb]
\includegraphics[scale=.265, clip=true, trim=8mm 8mm 2mm 37mm]{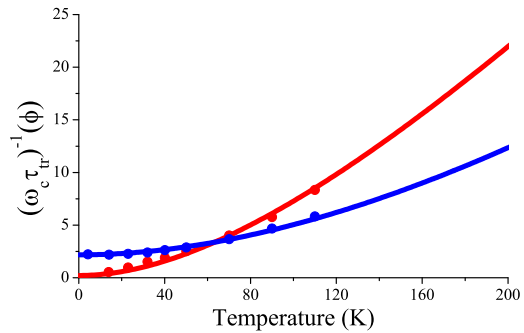}
\caption{ADMR $(\omega_{c} \tau_{tr})^{-1}$ versus T showing excellent agreement with HFL using \textit{no free parameters} (solid lines) and the data of ref. \cite{French2009} for OD $Tl_{2}Ba_{2}CuO_{6+\delta}$ (x=0.26). Shown are the isotropic contribution (Eq. (9a),blue) and the anisotropic term (Eq. (9b),red) that vanishes at the node. W$_{HFL}$ = 800K was used for the HFL bandwidth along ($\pi, \pi$), determined independently from both $\rho$(T) and $\cot \theta_{H}$, as discussed. To evaluate the pre-factor of Eqns. (9a) and (9b), we determined the nodal $\omega_{c}$ from $(\omega_{c} \tau_{H})^{-1}$ in Fig. 1 with $\tau_{H}^{-1} = T^{2}/W_{HFL}$. We account for the anisotropic $k_{F}$ with $k_{F}(\pi,0) / k_{F}(\pi,\pi)$=9/10 (ref. \cite{Plate2005}) and anisotropic $v_{F}$ with $v_{F}(\pi,0) / v_{F}(\pi,\pi)$=1/2 (refs. \cite{Plate2005, Analytis2007}). The T=0, impurity offset in the blue curve is arbitrary as it is sample specific. (Color online)}
\end{figure}

We now extend the ``bottleneck'' form for $\tau_{tr}$, connecting the projected space with the HFL, to include the anisotropic $k_{F}(\phi)$ and $v_{F}(\phi)$ ($\phi$ is the angle with respect to (0,0) - ($\pi$,0) in the Brillouin zone). The inverse HFL lifetime, $\tau_{HFL}^{-1}(\phi) =T^{2}/W_{HFL}(\phi)$, will share the anisotropy of $(k_{F}^{2}(\phi)/m^{*}(\phi))^{-1} =(v_{F}(\phi)k_{F}(\phi))^{-1}$ (recall W$_{HFL} \approx \epsilon_{F} g_{t}^{2}(x)$, which would essentially be an equality if not for J). The anisotropic $k_{F}$ and $v_{F}$ will also be evident in $\omega_{c}(\phi) = eBv_{F}(\phi)/ \hbar k_{F}(\phi)$. Evaluating the quantity measured in the ADMR\cite{Hussey2003, Abdel-Jawad2006, Analytis2007, French2009} experiments,
\begin{equation}
\displaystyle \frac{1}{\omega_{c} \tau_{tr}} (\phi) = \frac{\hbar k_{F}}{v_{F} e B \tau_{tr}} + \frac{1}{\omega_{c} \tau_{res}}
\end{equation}
\begin{center}
{\scriptsize $\displaystyle = \left[ \frac{2\pi p \hbar}{eB} \right] \left[ \frac{k_{F}(\phi)}{v_{F}(\phi)} \right] \frac{T^{2}}{T + 2\pi p W_{HFL}^{(\pi, \pi)} \left[ \frac{k_{F}(\phi)}{k_{F}^{(\pi, \pi)}} \right] \left[ \frac{v_{F}(\phi)}{v_{F}^{(\pi, \pi)}} \right]} + \frac{\hbar k_{F}(\phi)}{eB \lambda_{res}(\phi)}$}
\end{center}
where $\tau_{res}$ and $\lambda_{res}$ are the T=0, residual lifetime and mean-free-path. The residual term will exhibit the slight anisotropy of $k_{F}/ \lambda_{res}$, for simplicity we'll assume it's isotropic and discuss this subtle point elsewhere. Along ($\pi, \pi$), ADMR exhibits only the isotropic contribution. This isotropic term, Eq. (9a), corresponds to Eq. (8) with $k_{F}(\phi) = k_{F}(\pi, \pi)$ and $v_{F}(\phi) = v_{F}(\pi, \pi)$. Along $(\pi,0)$, both the isotropic and anisotropic contributions are present - corresponding to Eq. (8) with $k_{F}(\phi)=k_{F}(\pi, 0)$ and $v_{F}(\phi)=v_{F}(\pi, 0)$.  The anisotropic term, Eq. (9b), measured in ADMR is just the difference of Eq. (8) evaluated at ($\pi, 0$) and ($\pi, \pi$). This subtraction yields an anisotropic term that appears somewhat T-linear, interpreted in the experimental papers\cite{Abdel-Jawad2006, French2009} as having a physically significant T-linear contribution, even as T $\rightarrow$ 0. The physical motivation for this contribution, however, was left as an open question. Decomposing the data into these two terms is simply a convenient manner to compare our theory with the data - we stress that there is no physical significance of the anisotropic term.

Generally speaking, since $k_{F}(\pi, 0)/k_{F}(\pi, \pi)$ and $v_{F}(\pi, 0)/v_{F}(\pi, \pi)$ are less than unity, HFL predicts that $(\omega_{c} \tau_{tr})^{-1}$ is more linear (W$_{HFL}(\phi)$ decreases) and increases in magnitude off the node. Further consistent with the experiment, both T and T$^{2}$ terms are present at all angles about the Fermi surface. In Fig. 2, we find excellent agreement with the ADMR data\cite{French2009} that extends to the highest temperature currently available of 110 K - we use Eqns. (9a) and (9b) with \textit{no free parameters}.
{\scriptsize
\begin{subequations}
\begin{eqnarray}
&\displaystyle \frac{1}{(\omega_{c} \tau_{tr})_{iso}} = \left[ \frac{2 \pi p \hbar}{eB} \right] \left[ \frac{k_{F}^{(\pi , \pi)}}{v_{F}^{(\pi , \pi)}} \right] \frac{T^{2}}{T + 2 \pi p W_{HFL}^{(\pi , \pi)}} + \frac{\hbar k_{F}^{(\pi , \pi)}}{eB \lambda_{res}^{(\pi , \pi)}}& \\
&\displaystyle \frac{1}{(\omega_{c} \tau_{tr})_{aniso}} = \frac{1}{\omega_{c}^{(\pi , 0)} \tau_{tr}^{(\pi , 0)}} - \frac{1}{(\omega_{c} \tau_{tr})_{iso}}& \\
&\displaystyle = \left[ \frac{2 \pi p \hbar}{eB} \right]  \biggl( \left[ \frac{k_{F}^{(\pi, 0)}}{v_{F}^{(\pi, 0)}} \right] \frac{T^{2}}{T + 2 \pi p W_{HFL}^{(\pi, \pi)} \left[ \frac{k_{F}^{(\pi, 0)}}{k_{F}^{(\pi, \pi)}} \right] \left[ \frac{v_{F}^{(\pi, 0)}}{v_{F}^{(\pi, \pi)}} \right]}& \nonumber \\
&\displaystyle - \left[ \frac{k_{F}^{(\pi, \pi)}}{v_{F}^{(\pi, \pi)}} \right] \frac{T^{2}}{T + 2 \pi p W_{HFL}^{(\pi, \pi)}} \biggr)& \nonumber
\end{eqnarray}
\end{subequations}
}
At optimal doping, focusing on $Bi_{2}Sr_{2-x}La_{x}CuO_{6+\delta}$, $\rho$(T) is more linear than expected - W$_{HFL}$ is about half that deduced from $\cot \theta_{H}$.\cite{Anderson2009} But we believe the 60 T field applied\cite{Ono2000} is insufficient to suppress the fluctuation conductivity and $\rho$(T) is very sensitive to any vortex liquid\cite{Anderson2006b} contribution acting to linearize it. A precise measure of the resistive H$_{c2}$ is hard and any measure of H$_{c2}$ inevitably depends on somewhat arbitrary criteria. But H$_{c2}$ extrapolated from Nernst is much higher than traditionally thought - on the order of 60 T in optimally doped $Bi_{2}Sr_{2-x}La_{x}CuO_{6+\delta}$ and much higher in other cuprates.\cite{Wang2003} Larger fields are needed to access the true field-induced normal state at low T near optimal doping.

However, we did find self-consistency of the nodal laser-ARPES $(k-k_{F})^{2}$ scattering rate and $\cot \theta_{H}$ at optimal doping.\cite{Anderson2009} So given the self-consistent W$_{HFL}$=800K for this OD sample, we make a well-defined prediction for the laser-ARPES scattering rate. Using the projective line shape (Eq. (1) of ref. \cite{Casey2008}) to fit the nodal EDC's (or MDC's but only for $|v_{F} (k-k_{F})| \lesssim W_{HFL}$), should yield Eq. (5a) with $\pi$p=0.43 and C$_{HFL}$=2.2eV$^{-1}$. The linear-T term should be quite isotropic but the anisotropic $v_{F}$ and $k_{F}$ will increase C$_{HFL}$ off the node, as discussed C$_{HFL} \propto (k_{F}(\phi)v_{F}(\phi))^{-1}$. The predicted T-linear pre-factor of 0.43 may differ slightly than measured as we found at optimal doping,\cite{Casey2008} in as much as the Tt $\gg$ 1 requirement of the finite temperature extension\cite{Yuval1970} is upheld and the slight experimental resolution is included.

Within HFL, expressions were derived (Eq. (5)-(7),(9); refs. \cite{Casey2008, Anderson2006} for IR conductivity; ref. \cite{Casey2010} for entropy transport) relating the normal state transport and spectroscopic properties with self-consistency in OD $Tl_{2}Ba_{2}CuO_{6+\delta}$. The framework is amply detailed to guide further experimental confirmation in other cuprates and dopings, particularly for laser-ARPES, ADMR, and $\rho$(T$\rightarrow$0) in magnetic fields exceeding H$_{c2}$.
\bibliography{PRLCaseyAnderson}
\end{document}